\documentclass[twocolumn,showpacs,superscriptaddress,preprintnumbers,floats,aps,pra, nofootinbib,10pt]{revtex4-1}
\usepackage{epstopdf}
\usepackage{ulem}
\usepackage{amssymb}
\usepackage{natbib}
\usepackage{times}
\usepackage{graphicx,bm,amssymb}
\usepackage{pstricks}
\usepackage{psfrag}
\usepackage[colorlinks=true,linkcolor=blue,citecolor=blue]{hyperref}
\newcommand{\be}{\begin{equation}}
\newcommand{\ee}{\end{equation}}
\newcommand{\bse}{\begin{subequations}}
\newcommand{\ese}{\end{subequations}}
\newcommand{\bary}{\begin{eqnarray}}
\newcommand{\eary}{\end{eqnarray}}

\usepackage{graphicx}
\newcommand{\ergs}{\,erg\,s$^{-1}$}

\newcommand{\doce}{\times10^{12}}
\newcommand{\once}{\times10^{11}}
\newcommand{\trece}{\times10^{13}}

\newcommand{\ttres}{\times10^{33}}

\newcommand{\one}{XMMU J173203.3-344518\,\,}

\newcommand{\five}{1E 1207.4-5209\,\,}
\newcommand{\six}{CXOU J160103-513353\,\,}
\newcommand{\seven}{1WGA J1713.4-3949\,\,}
\newcommand{\eight}{XMMU J172054.5-372652\,\,}
\newcommand{\nine}{CXOU J085210.4-461753\,\,}

\newcommand{\teff}{T_{\textup{  eff}}^\infty}

\def\aj{AJ}
\def\apj{ApJ}
\def\apjl{ApJ}
\def\apjs{ApJS}
\def\aap{A\&A}
\def\mnras{MNRAS}
\def\prd{Phys.~Rev.~D}
\def\nat{Nature}
\def\physrep{Phys.~Rep.}

\begin{document}
\title{A hypercritical accretion scenario in Central Compact Objects\\   accompanied with an expected neutrino burst}
             
\author{N. Fraija} 
\altaffiliation{nifraija@astro.unam.mx}
\affiliation{Instituto de Astronom\' ia, Universidad Nacional Aut\'onoma de M\'exico, Circuito Exterior, C.U.,\\ A. Postal 70-264, 04510 M\'exico D.F., M\'exico.}

\author{C. G. Bernal}
\affiliation{ Instituto de Matem\'atica, Estat\'istica e F\'isica, Universidade Federal de Rio Grande,  Av. Italia\\ km 8 Bairro Carreiros, Rio Grande, RS, Brazil.}

\author{G. Morales}
\affiliation{Instituto de Astronom\' ia, Universidad Nacional Aut\'onoma de M\'exico, Circuito Exterior, C.U.,\\ A. Postal 70-264, 04510 M\'exico D.F., M\'exico.}

\author{R. Negreiros}
\affiliation{Instituto de F\'isica, Universidade Federal Fluminense, Av. Gal. Milton Tavares de Souza s/n,\\ Gragoata, Niteroi, 24210-346, Brazil.}


\begin{abstract}
The measurement of the period and period derivative, and the canonical model of dipole radiation have provided a method to estimate the low superficial magnetic fields in the so-called Central Compact Objects (CCOs). In the present work, a scenario is introduced in order to explain the magnetic behavior of such CCOs. Based on magnetohydrodynamic simulations of the post core-collapse supernova phase during the hypercritical accretion episode, we argue that the magnetic field of a newborn neutron star could have been early buried. During this phase, thermal neutrinos are created mainly by the pair annihilation, plasmon decay, photo-neutrino emission and other processes.  We study the dynamics of these neutrinos in this environment and also estimate the number expected of the neutrino events with their flavor ratios on Earth.  The neutrino burst is the only viable observable that could provide compelling evidence of  the hypercritical phase and therefore, the hidden magnetic field mechanism as the most favorable scenario to explain the anomalous low magnetic fields estimated for CCOs.
\end{abstract}

\pacs{14.60.Pq, 97.60.Jd}
\maketitle
\section{Introduction}\label{sec-Intro}
After the launch of the \textit{Chandra X-ray Observatory} (hereafter, \textit{Chandra}), substantial observations have been performed in the well-known Central Compact Objects (CCOs). The most relevant characteristics of these sources have been widely studied, although only a few works have been performed  (see, \citep{2013ATel.5046....1G} and the references within it). Currently, there are nine CCOs confirmed with common features \citep{2015ApJ...808..130L}: i) The location of these compact objects has been found near the center of some young supernova remnants  (SNRs, \citep{2010PNAS..107.7147K}); ii)  A counterpart emission detected in optical or radio wavelengths; iii) The non detection of pulsar wind nebulae from these sources and iv) The thermal spectrum observed in the soft X-ray band (peaking at $kT_{BB}\sim$ 0.2-0.6 keV) with high luminosities $L_X\sim 10^{33} - 10^{34}\,  {\rm erg\, s^{-1}}$ \citep{2004IAUS..218..239P, 2008AIPC..983..311D, 2015A&A...573A..53K}. \\
 Although the superficial magnetic field in CCOs has not been observed,  calculations of this one have been found atypically low, favoring the hidden magnetic field scenario. In this model, during a core-collapse supernova the high strength of  magnetic field is submerged by the hypercritical accretion phase\footnote{The hypercritical accretion phase occurs when the accretion rate is as high as the Eddington accretion rate ($\dot{m}>\dot{m}_{\text{Edd}}$).} (see e.g., \cite{1995ApJ...440L..77M}, \cite{1999A26A...345..847G}, \cite{Bernal2013} and references therein).  Muslimov and Page \cite{1995ApJ...440L..77M} investigated the extreme conditions inside core-collapse supernova. These are the convective envelope, the hyper-accretion of material and the submergence of the magnetic field on the stellar crust. Using these conditions, simple 1D ideal magnetohydrodynamic (MHD) simulations were performed to show that the hypercritical accretion is related with the submergence of the magnetic field \citep{1999A26A...345..847G}. These simulations showed a prompt submergence of the bulk magnetic field into the neutron star (NS) crust. Later, several authors have revisited this scenario in order to study the magnetic field evolution in the CCOs \citep{2011MNRAS.414.2567H, 2012MNRAS.425.2487V, Bernal2013, 2015MNRAS.451..455F, 2016MNRAS.462.3646B}.  Bernal et al. \cite{Bernal2013} used high refinement of  MHD numerical simulations confirming, in general,  the submergence of the magnetic field given by the hypercritical accretion.  Fraija and Bernal \cite{2015MNRAS.451..455F}  showed that thermal neutrinos play an outstanding role  in the formation of a quasi-hydrostatic envelope. They showed that resonant conversions of neutrino flavors are expected during the hypercritical accretion phase.    Authors applied the works performed in \cite{Bernal2013} to Cassiopeia A and studied the dynamics of the propagation and neutrino oscillations in the newly forming neutron star crust. In addition, they estimated the number of neutrino events that can be generated from the hypercritical phase and be expected on Earth.  Bernal and Fraija  \cite{2016MNRAS.462.3646B} included a new and more adequate customized EoS routine for the extreme conditions presented in the system as well as the radiative cooling by neutrinos from various processes. The customized EoS routine was applied to Kesteven 79.\\
In the present work, we use the \textsc{flash} code in order to simulate the hypercritical phase into the stellar surface, including the magnetic field evolution, the magnetic reconnections processes and the neutrino luminosity for the known CCOs. In addition, the thermal neutrino flux with its flavor ratio expected on Earth is computed. The paper is arranged as follows. In section \ref{sec-CCOs} we describe the main characteristics of CCOs. In section \ref{sec-Physics} we describe the physics associated to the hypercritical model including the magnetic field submergence and reconnections. In section \ref{sec-Neutrinos} we study the thermal neutrino dynamics,  analyze the neutrino oscillations and estimate  the number expected of the neutrino events with their flavor ratios on Earth. Finally, the conclusions are shown in Section \ref{sec-Results}. 
\vspace{0.8cm}
\section{Central Compact Objects}\label{sec-CCOs}
CCOs are soft X-ray sources located inside SNRs. It is amply accepted that these compact objects are young and radio-quiet Isolated Neutron Stars (INSs). As follows, we mention the most relevant  observable quantities which are summarized in Table \ref{Table:CCOs}.
\subsection{\one in G353.6-0.7}
The CCO XMMU J173203.3-344518 was discovered by \textit{XMM-Newton} in 2007 as a point-like X-ray source roughly at the center of the TeV-emitting SNR HESS J1731-347 (a.k.a. G 353.6-0.7). Taking into consideration the Sedov solution for the SNR, Tian et al. \cite{2008ApJ...679L..85T} estimated the age to be $27$ kyr for a distance of $d=3.2$ kpc. Using models based on the carbon atmosphere spectra at distances of $d=3.2$ kpc and $d=4.5$ kpc, Klochkov et al. \cite{2015A&A...573A..53K} predicted effective temperatures of ${\small \teff}=1.78_{-0.02}^{+0.04}$ K and $1.82_{-0.01}^{0.03}$ K,  respectively. Considering the previous values of temperatures and distances, values of magnetic fields in the range of  $B\gtrsim(10^{10}-10^{11})$ G and $\sim10^{12}$ G were suggested. The best-fit curve of the energy spectrum was a black-body (BB) function with $kT\sim0.5$ keV and $k$ the Boltzmann constant  \citep{acero2009x}. 
\subsection{\five  in PKS 1209-51/52}
The object  \five located at a distance of $d\sim2.1$  kpc was discovered by the \textit{Einstein} satellite \citep{1984AJ.....89..819H} at $\sim$ 6' from the center of the SNR G296.5+10.0 \citep{2000AJ....119..281G}. Analyzing the radio and the optical observations from this CCO, an age of $\sim7$ kyr was estimated in Ref. \cite{1988ApJ...332..940R}. Pulsations from this CCO were detected with a period of $P=0.424$ s \citep{2000ApJ...540L..25Z} and a period derivative of $\dot{P}=2.0_{-1.3}^{+1.1}$ ss$^{-1}$ \citep{2002ApJ...574L..61S}. Using  the period and period derivative, Halpern and Gotthelf \cite{2011ApJ...733L..28H} constrained the value of the superficial magnetic field to be $B\lesssim9.9\times10^{ 10}$ G. In order to describe the X-ray emission of this CCO, two BB curves with parameters  $kT_1\sim0.16$ keV and $kT_2\sim0.3$ keV were used \citep{2004A&A...418..625D}.
\subsection{\six in G330.2+1.0}
SNR G330.2+1.0 was discovered in 2006 by the Advanced CCD Imaging Spectrometer (\textit{ACIS}) on board \textit{Chandra}. Particularly, the source \six located in the Southeast Hemisphere of the radio SNR center was the most interesting and brightest one. Park et al. \cite{2006ApJ...653L..37P} used a BB law with $T_{BB} =5.6_{-0.4}^{+0.3}$ K to fit the spectrum. Assuming the distance of $d=5$ kpc and three values of magnetic fields $B=(0,1,10)\trece $ G, Park et al. \cite{2009ApJ...695..431P} estimated that the  effective temperature was in the range of ${\small \teff} \sim(2.5-5.5)$ K. In addition,  using the Sedov solution with this distance, the age of this object was found to be  $\sim1.1$ kyr. The X-ray spectrum could be described by NS atmosphere models. Neither the radio nor the optical counterpart was found from this object. 
\subsection{\seven in G347.3-0.5}
The source \seven  was observed for the first time in 1992 by  \textit{ROSAT}   \citep{pfeffermann1996rosat} and re-observed with the Advanced Satellite for Cosmology and Astrophysics (\textit{ASCA})\citep{1999ApJ...525..357S}. This object located at the center of the SNR G347-3-0.5 was detected in the non-thermal X-ray band. This CCO presented neither optical counterpart nor radio emission. The best-fit curve for the X-ray spectrum was obtained with a BB function with parameter $kT\sim 0.4$ keV plus a power-law or the sum of two BB spectra (with $kT_1\sim 0.5$ keV and $kT_2\sim0.3$ keV). Using the Sedov solution for a SN with an initial energy of $\sim10^{51}$ erg at a distance of $d\sim 1.3$ kpc,  Cassam-Chenai et al.  \cite{2004A&A...427..199C} estimated an age of $t\sim2.1$ kyr. It is worth noting that the CCO associated with this SNR has been exhibiting  periods between 100 and 400 ms with an upper limit in the period derivative of $\dot{P}\geq 10$ ss$^{-1}$\citep{2008A&A...484..457M}.
\subsection{\eight in G350.1-0.3}
SNR G350.1-0.3 was observed in the X-ray band by \textit{ROSAT}  \citep{2008ApJ...680L..37G} and \textit{ASCA}. Inside SNR G350.1-0.3,  the source \eight stands out among all over the sources, being the brightest one. The best-fit curve to the spectrum follows a BB law with $kT\sim0.53$ keV for a distance of $d=4.5$ kpc and an age of $\sim0.9$ kyr. It is relevant to state that none pulsation longer than $6.4\ \text{s}$ has been found (see Ref. \citep{2011ApJ...731...70L}).
\subsection{\nine in G266.1-1.2 or ``Vela Jr.''}
This CCO, known as CXOU J085201.4-461753, was discovered by \textit{ROSAT}, at the south-east corner of the SNR Vela  between 1990/91 \citep{1998Natur.396..141A}  and re-observed with \textit{BeppoSAX} and \textit{Chandra} \citep{de2008central}. Using the Sedov approximation for a SNR in an adiabatic phase, the age was estimated at $\sim1$ kyr for a distance of $d\sim 1$ kpc. The best-fit model used to describe the X-ray spectrum was the BB function with $kT\sim 0.4 $ keV \citep{2002ApJ...580.1060K}. None evidence of X-ray pulsations or optical ($R-$band) counterpart has been found \citep{2007A&A...473..883M}.\\\\
\begin{table*}
\begin{center}
\caption{Observable quantities of the Central Compact Objects.} \label{Table:CCOs}
\label{t1}
\resizebox{12cm}{!}{
\begin{tabular}{lllllll}
\hline\hline
CCO                   & SNR            & $d$ (kpc) & $B$ (G) & $t$ (kyr) & $L_{\rm x}$ \ergs \\ \hline\hline
XMMU J173203.3-344518 & G 353.6-0.7    & 3.2 &  $(1-10)\once$   & 27    &      $1.1\ttres$          \\
1E 1207.4-5209        & PKS 1209-51/52 & 2.2    &   $9.8\times10^{10}$   &  7   &    $2.5\times10^{33}$ \\
CXOU J160103-513353   & G330.2+1.0 & 5.0  &   $(0,1,10)\doce$  & $\gtrsim3$  &$1.5\times10^{33}$           \\
1WGA J1713.4-3949     & G347.3-0.5     &  1.3   &     &   1.6 &        $3.8\times10^{33}$    \\
XMMU J172054.5-372652 & G350.1-0.3     &   4.5  &     &  $0.6-1.2$   &     $3.9\times10^{33}$        \\
CXOU J085201.4-461753 & G266.1-1.2     &    1.0 &     &   1  &       $2.5\times10^{33}$        \\
                      &                &     &     &     &            &     \\\hline\hline
\end{tabular}}
\end{center}
\end{table*}

%
%
\section{Numerical approach}\label{sec-Physics}
\begin{figure*}
\centering
\includegraphics[width=0.8\textwidth]{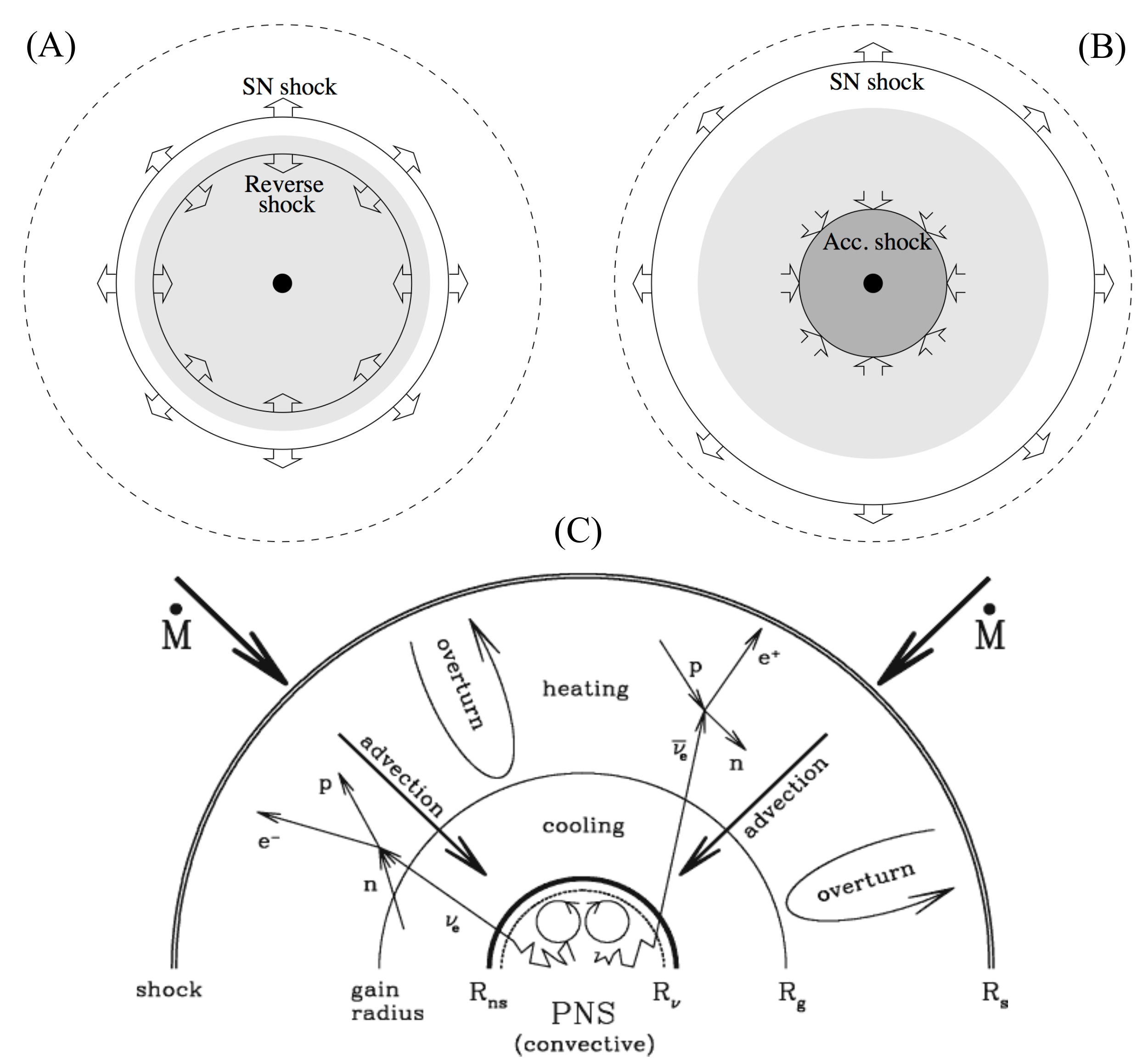}
\caption{Hypercritical accretion phase schematic. (A) The formation of reverse shock; (B) the formation of a quasi-hydrostatic envelope with an accretion shock and (C) the complex dynamics around the newborn NS during such phase.}
\label{fig:1}
\end{figure*}

Considering these CCOs as newborn NSs, we analyze the hypercritical accretion phase inside a core-collapse supernova, focusing on the dynamics and morphology of the magnetized and the thermal plasma around the proto-NS. We use the model previously shown in Ref. \cite{Bernal2013}.  Here, we are interested in following the early evolution of the magnetic field when the hypercritical accretion phase occurs. This phase remains active while the reverse shock reaches the NS surface seconds after the core-collapse. Figure \ref{fig:1} shows a schematic view of the different phases: i) the  reverse shock, ii) the formation of a quasi-hydrostatic envelope and iii) the complex dynamics around the newborn NS.  In addition, the neutrino cooling processes have been included.\\
In order to tackle numerically the hypercritical accretion phase, a customized version of the Eulerian and multi-physics parallel \textsc{flash} code is used \citep{Fryxell2000}. The complete set of MHD equations is solved using  the unsplit staggered mesh algorithm which is included in the most recent \textsc{flash} version code \footnote{\url{http://flash.uchicago.edu/site/}} (\textsc{flash} 4).  The algorithm is based on a finite-volume, high-order Godunov method combined with a constrained transport (CT) type of scheme which ensures the solenoidal constraint of the magnetic fields on a staggered mesh geometry. In this approach, the cell-centered variables such as the plasma mass density, the plasma momentum density and the total plasma energy are updated via a second-order MUSCL-Hancock unsplit space-time integrator using the high-order Godunov fluxes. The rest of the cell face-centered (staggered) magnetic field is updated using Stokes' Theorem as applied to a set of induction equations, enforcing the divergence-free constraint of the magnetic fields. In this way, the USM scheme includes multidimensional MHD terms in both normal and transverse directions, satisfying a perfect balance law for the terms proportional to the magnetic field divergence in the induction equations \citep{LeeDongwook2013}.
\\
Regarding the thermodynamical variables, the EoS algorithm implements the equation of state needed by the MHD solver. This  EoS function  provides to the interface for operating on a one-dimensional vector. The same interface can be used for a single cell by reducing the vector size to unity. Additionally, this function can be used to find  thermodynamic quantities either from the density, temperature, and composition or from the density, internal energy, and composition. In the present work, we use the Helmholtz package, which in turn uses a fast Helmholtz free-energy table interpolation to handle degenerate/relativistic electrons/positrons and which also incorporates the effects of radiation pressure and ions (via the perfect gas approximation). It includes contributions coming from radiation, ionized nuclei, and degenerate/relativistic electrons and positrons. The pressure and the internal energy are calculated as the sum over the components.   The Helmholtz EoS has been widely tested in several astrophysical environments (i. e.  core-collapse modeling), but not explicitly in this kind of applications. Therefore,  it will be used in the work with caution. Full details of the Helmholtz algorithms are provided in \cite{Timmes2000}.\\
The neutrino cooling processes are also included in this code through a customized Unit  and it will be discussed along  the following section.\\
\subsection{Initial and boundary conditions}
We perform 2-dimensional simulations during the hypercritical phase taking into consideration the domain covered by the region $2 \times 10^6$ cm at the base and $10^7$ cm at the height for a typical NS radius  of 10 km and mass of 1.45 $\mathrm{M}_{\odot}$. The mesh resolution for our simulations was  512 $\times$ 5120 effective zones. The refinement criterion used by the \textsc{flash} Code is adapted in Ref. \cite{LOHNER1989195}.  The Lhner's error estimator originally developed for finite element applications has the advantage of using a mostly local calculation. Furthermore, the estimator is dimensionless and can be applied with complete generality to any field variables of the simulation or any combination of them.
\\
The gravitational acceleration is taken as a plane-parallel external field  $g_y=-GM/y^2$.  The temperature is initially uniform with a value of $T=10^{9}$ K. The code finds the correct profile as well as the rest of the thermodynamic variables using the Helmholtz EoS.  
We choose as initial magnetic field configuration a loop shape because the curve geometry can play an important role in the dynamics of the plasma and in the magnetic reconnection processes close to the stellar surface. In the central part of the loop the B-field with a pulsar-typical value of $B_0 = 10^{12}$ G  extends in accordance with a Gaussian function given by $B(d) = B_0\times \exp\left(-{d/R_L}\right)^2$,  with $R_L = 1$ km and $d$ the distance to the column center. The two feet of the loop are centered at $x=-5$ km, and $x=+5$ km.\\
During the simulation, the accretion rate per area is considered to be constant for each CCO with values in the range of $\dot{m} = 10^{13} -10^{15} \; \mathrm{g \, cm^{-2} \, s^{-1}}$,  which are typical for supernova progenitor models.
The simulation starts when the flow, driven by the reverse shock, falls freely on the stellar surface with velocity and density profiles given by
\begin{equation}
v_\mathrm{ff} = \sqrt{2GM/y}  
\;\;\;\; \mathrm{and} \;\;\;\;
\rho_\mathrm{ff} = \dot{m}/v_\mathrm{ff}\,,
\label{Eq:free-fall}
\end{equation}
respectively.  For the boundary conditions, we consider at the top of the computational domain the velocity profile ($v_\mathrm{ff}$) of the non-magnetized matter in free fall, and set a density profile ($\rho_\mathrm{ff}$)  which corresponds to the desired  accretion rate. At the bottom, on the NS surface, we use a custom boundary condition that enforces the hydrostatic equilibrium. In addition, we implement periodic conditions in  lateral directions. Periodic (wrap-around) boundary conditions are initially configured in this routine as well.\\
For the magnetic field, the two lateral sides are also treated as periodic boundaries, while at the bottom the field is frozen from the initial condition. Nuclear reactions are not considered here.
We use the Courant number of  0.5 resulting in a typical time step of  $10^{-6}$ s. 
\subsection{Magnetic field Submergence and Reconnections}
\begin{figure*} [!htb]
   \centering
   \includegraphics [width=0.8\textwidth]{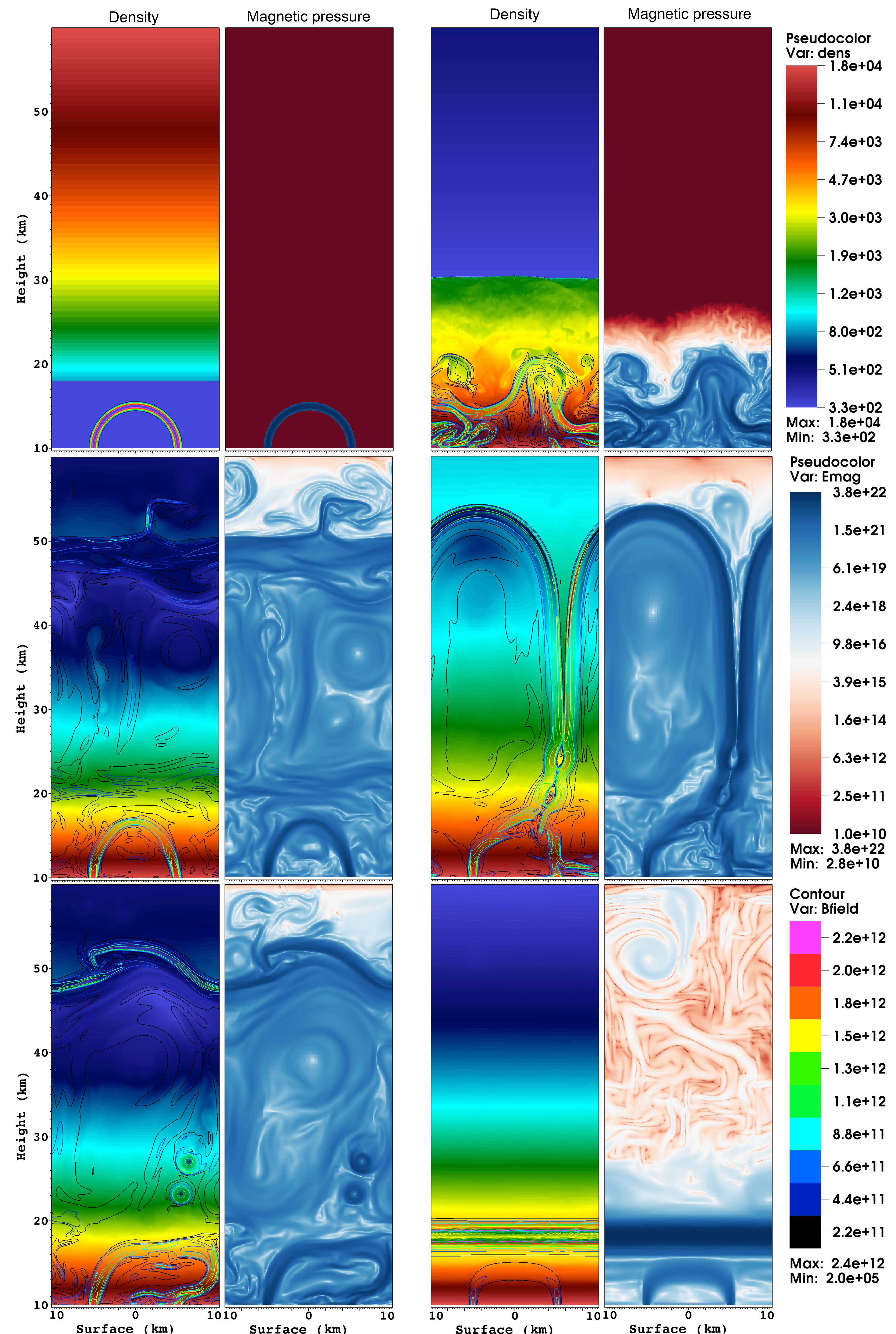}
   \caption{Color maps of density, with magnetic field contours superimposed, and magnetic energy for the case $\dot m=0.001 \dot m_{0}$. We show various time-steps of the evolution of the magnetic loop to analyze the magnetic reconnection process. Upper panels: (left: $t=0$ ms; right: $t=1$ ms); Middle panels: (left: $t=20$ ms; right: $t=100$ ms); Bottom panels: (left: $t=400$ ms; right: $t=1000$ ms).}
   \label{fig:2}
\end{figure*}

\begin{figure}
   \centering
   \includegraphics [width=0.5\textwidth]{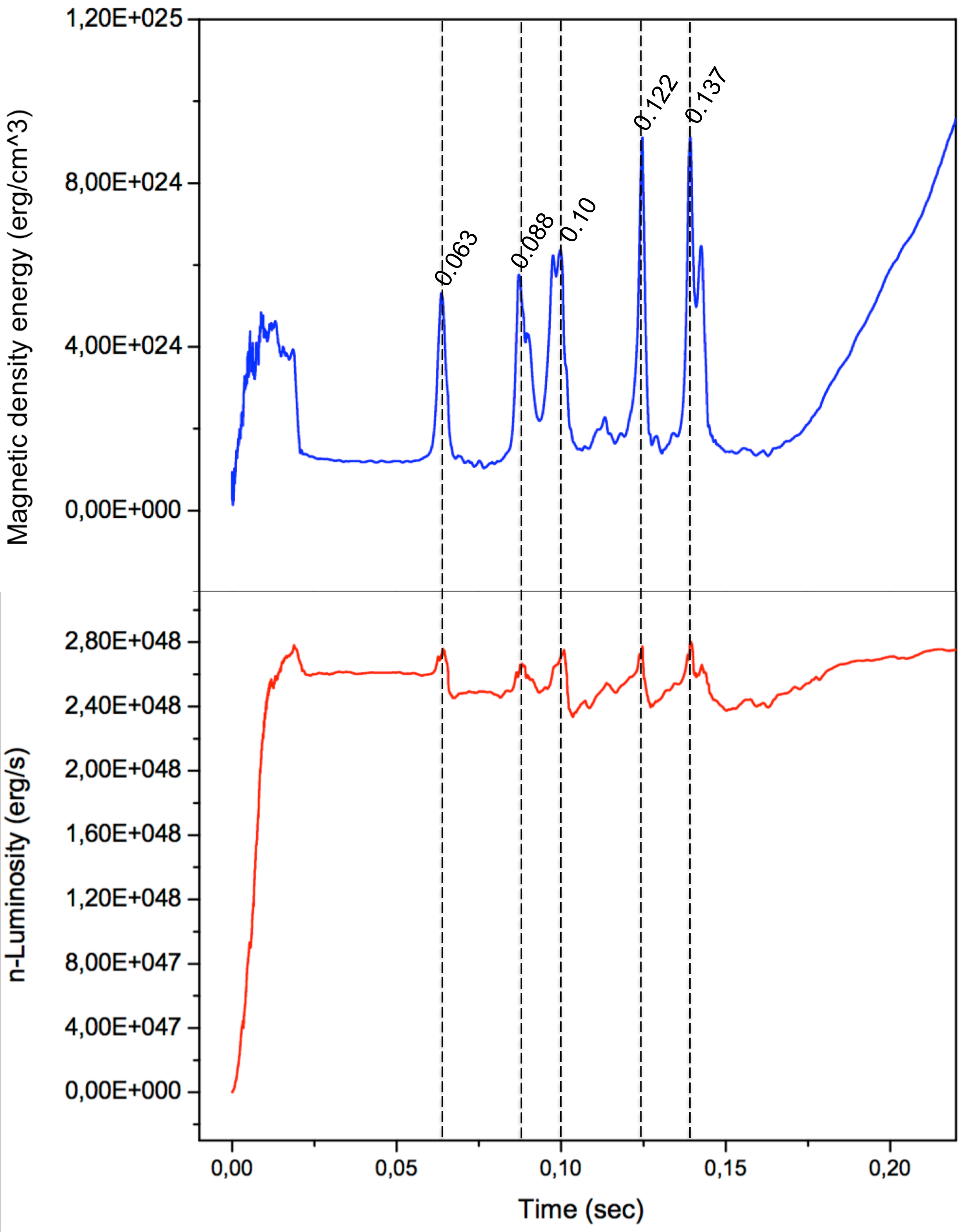}
   \caption{Time evolution of magnetic energy density integrated in whole computational domain and neutrino luminosity due different cooling processes included in the model, as were described in \citep{Itoh1996}. The peaks correspond to various reconnection episodes presents in the system. Note small peaks as contributions from magnetic reconnection.}
   \label{fig:4}
\end{figure}

As already shown in Ref.  \cite{Bernal2013}, for accretion rates higher than few thousands of  the Eddington accretion rate, the stellar magnetic field is submerged into the crust of the newly born NS. Considering CCOs as typical young NSs, we perform numerical simulations of the hypercritical phase for a range of accretion rates, founding that the submergence of the magnetic field is absolute and that the submergence timescale only depends on the hypercritical accretion rate, for models of typical NS.  Here, we want to show a model of weak hyper-accretion which allows the dynamic behavior of the system to be analyzed  in detail.  The interpretation and discussion of the results are given as follows. Figure \ref{fig:2} shows the temporal evolution of matter density superimposed with the magnetic field contours, and also the magnetic energy density for a fiducial low accretion rate of $\dot{m} = 1.6 \times 10^{13} \; \mathrm{g \, cm^{-2} \, s^{-1}}$ (roughly 50 $\mathrm {{M}_{\odot} \, yr^{-1}}$ ). The top panels exhibit both the initial state (left; $t=0$ ms)  with the reverse shock before it reaches the magnetic field loop and the initial strong transient phase (right; $t=1$ ms) with a changed magnetic loop topology when the reverse shock compresses the plasma on the NS surface.      During this phase, the expansive shock interacts with the flow in free fall. The middle panels display the complex dynamics close to the NS surface (left; $t=20$ ms), including the original loop restructured due to  reconnection processes. When the flow interacts with the NS surface and rebounds, a portion of the magnetic field is entrained out forming a magnetic layer above the loop (right; $t=100$ ms). The magnetic Rayleigh-Taylor instabilities in the computational domain induce magnetic reconnections between the loop and layer. The non-magnetized flow falling on the stellar surface encounters in its way the magnetic layer. These interactions induce the formation of  large and isolated fingers which in turn reach the magnetic loop, thus producing  reconnection processes several times. The lower panels show the phase of quasi-hydrostatic equilibrium between the magnetic layer and the falling flow (left; $t=400$ ms). After several rebounds of the falling flow,  the fingers have disappeared and the magnetic loop suffers small deformations up to it returns to the original shape. It is worth noting that during this phase, the  matter pressure begins to be dominant. The final state is reached (right; $t=1000$ ms). This panel shows that the quasi-hydrostatic envelope is well-established and also exhibits the magnetic loop resisting to the hyper-accretion. During this moment, most of the MHD instabilities have disappeared in the system. The magnetic loop seems to be more flattened due to the high accretion over it. Finally, the magnetic field is confined into the new crust, although residual magnetic fields with small strengths  are within the quasi-hydrostatic envelope without affecting the dynamics.\\
The adiabatic and radiative gradients can be estimated through simulations when the system has reached the quasi-hydrostatic equilibrium. The adiabatic gradient is $\nabla _{ad}=1-1/\gamma _{c}\simeq 0.27$ and the radiative gradient is $\nabla _{rad}=\left( d\ln T/d\ln P\right) \simeq 0.26$. Note that the value of the adiabatic index $\gamma _{c}=1.35$ has been taken directly from the simulation, and the radiative gradient was calculated using the temperature and pressure profiles. It is worth noting that  the gradients are almost constant within the envelope, except in the region near the NS surface. Due $\nabla _{rad}\lesssim \nabla _{ad}$, the system is manifestly stable to convection, although these values close enough numerically would  indicate probably a marginal stability. Within the envelope the flow is fully subsonic, as expected after passing through the accretion shock front (the sound speed is $c_{s}=\sqrt{\gamma _{c}P/\rho }\simeq 7.3\times 10^{9}\,\mathrm{cm}\ \mathrm{s}^{-1}$ and $v\simeq 1.3\times 10^{8}\,\mathrm{cm}\,\mathrm{s}^{-1}$ for a Mach number of $m=v/c_{s}\simeq 10^{-2}$).  Therefore, the global structure of the accretion column can be studied in detail and compared with the analytical approach \citep{1989ApJ...346..847C}, particularly in the region where the approximations in the latter break down.\\     
In the final stage of the simulations, when the system is relaxed hundreds of milliseconds after the reverse shock rebounds on the stellar surface, an accretion envelope is established which is in quasi-hydrostatic equilibrium and separated from the continuous accretion inflow by an accretion shock. The material piled up by hyper-accretion builds a new stellar crust, submerging the magnetic field.  From the simulations, it is inferred that at the base of the envelope the magnetic field is confined and amplified inside the crust with strengths ranging between $\sim 10^{12}$ and $10^{13}$ G.   Figure \ref{fig:3} shows the radial profiles of density, pressure, and velocity obtained from the simulations and compared with the analytical solutions described in Ref. \cite{Chevalier2005}. Note that the density profile displays the accreted material in the new crust (top panel). This material increases the pressure at the base of envelope (middle panel).   Due to the additional degrees of freedom in the simulations, the fluid can flow freely through the lateral sides, as shown in the bottom panel. This includes both the radial and horizontal velocities.   It is important to highlight that the thermodynamical conditions in the fluid depend on the accretion towards the proto-NS. The Fermi temperature can be computed from the Fermi energy $E_{F}$ as
\begin{equation}
T_{F}=\sqrt{p_{F}^{2}+m_{e}^{2}} -m_{e}\simeq 6.48\times 10^{10}\,\mathrm{K,}
\end{equation}
at the base of the flow, where $p_{F}=(3\pi ^{2}n_{e})^{1/3} $ is the Fermi momentum, $m_e$ is the electron mass.  Hereafter, the natural units $\hbar=c=k=1$  will be used.  The temperature obtained from the simulation, close to the bottom of the accretion column in quasi-stationary state is $T\simeq 4.62\times 10^{10}\,\mathrm{K}$, and then,  $T/T_{F}\simeq 0.7$.  It is thus clear that the assumption of the $e^{\pm}$ degeneracy is not a proper approximation, and a full expression such as the one in the Helmholtz equation of state is required if one wishes to compute the evolution of the flow accurately. It is also clear that neutrino cooling effectively turns on a scale height $r\simeq4\times10^{5}\:\mathrm{cm}$.   For the current simulation the mean value of emissivity in such region is $\dot{\epsilon}_{\nu}\simeq1.6\times10^{30}\:\mathrm{erg\, s^{-1}\, cm^{-3}}$. Therefore, the integrated neutrino luminosity is given by $L_{\nu}\simeq2.6\times10^{48}\:\mathrm{erg\, s^{-1}}$.\\
\noindent The top panel in Figure  \ref{fig:4} shows the magnetic  energy density integrated into the whole computational domain. Several significant peaks are observed on the energy curve.  Each peak corresponds to different magnetic reconnection episodes suffered by the loop (due to interactions between the magnetic Rayleigh-Taylor instabilities and the loop).  Magnetic energy is converted into thermal energy by the code. Magnetic reconnection  with very strong magnetic fields, as the current case, is a problem poorly understood to date, hence there is not a robust theory that can sustain the results. The bottom panel in Figure  \ref{fig:4} shows the total luminosity due to several neutrino processes present in the neutrino-sphere of the newly born NS integrated in the whole computational domain. After the initial transient phase, the neutrino luminosity has small oscillations around a fixed value, with some small peaks, corresponding to the magnetic reconnection episodes suffered by the loop. \\
In this phase, the neutrino loss rate, the temperature and the density near the NS are steadily increased in the new crust. Eventually, neutrino losses are expected to initiate the collapse of the envelope. This may reduce the pressure in the envelope so that the effect of the neutrino losses is reduced.

Once emitted, neutrinos inside the optically thin material will act like an energy sink.  Under the present conditions of density and temperature at the base of the flow, the fluid consists mainly of free neutrons, protons and electrons. Here, we are interested in following the evolution of the initial transient phase when a portion of the magnetic field, confined within the new NS crust at the base of the envelope, can escape away from the surface.
As noted by Bernal et al. \cite{Bernal2013}, the crust plays an important role in the NS evolution and its dynamics because it is an interface which separates the NS interior from the external environment. In the present work, electrical resistivity of the crust is expected to be important in the evolution of the submerged magnetic field.\\ 
\noindent When the shock stabilizes the magnetic field, it suffers a complete submergence in a scale of few hundreds of meters above the stellar surface, and  four dynamically distinct regions can be identified in the pre-supernova. Table \ref{table1} displays a brief description including the densities and radii of each region. \\
Finally, we are now in the position to analyze the dynamics of the quasi-hydrostatic envelope, in the phase when the hypercritical regime stops. Since for large  hypercritical accretion rates (roughly 500 - 1000 $\mathrm {{M}_{\odot} \, yr^{-1}}$ ) the submergence of the magnetic field is more efficient (and faster), we consider this hyper-accretion to analyze the re-diffusion of a portion of the magnetic field from the new NS crust.  We start the simulations considering as initial conditions the final state when a quasi-hydrostatic envelope is formed around the newborn NS, including the new crust with the submerged magnetic field. The radial profiles of the envelope can be modeled as shown in \cite{1989ApJ...346..847C}.   The new crust formed in the hyper-accretion phase lies at the base of such envelope with the magnetic field confined in it.  We want to model the crust structure assuming  a power-law radial profile for density $\rho=\rho_{e}(\delta y/y)^n$ and pressure $p=p_{e}(\delta y/y)^m$, where $\rho_{e}$ and $p_{e}$ are the values of the density and pressure at the base of the envelope, respectively. The power indexes  $n$ and $m$ are chosen such that they match with the results of previous simulations. The radial velocity profile in the crust is assumed to be equal zero and the magnetic field is modeled as it was horizontally confined inside the crust, with strengths ranging between $\sim 10^{12}$  and $10^{13}$ G. In this case, it is not allowed  accretion through the outer boundary because at this stage the hyper-accretion is over. However, this boundary (outflow-type) allows the material to leave the computational domain due to the zero-gradient boundary condition. Although this type of boundary could appear natural for the current problem, it is likely incorrect as soon as the flow reaches the boundary (see \cite{2005ApJ...625..347G}). These authors considered this problem in the framework of classical nova modeling.  It is worth noting that it should be addressed for future works.    In the second set of simulations, periodic lateral conditions are imposed for the matter and the magnetic field, as the hypercritical simulations shown above.   The NS surface was simulated as a hydrostatic boundary condition.

\begin{figure*} [!htb]
   \centering
   \includegraphics [width=0.9\textwidth]{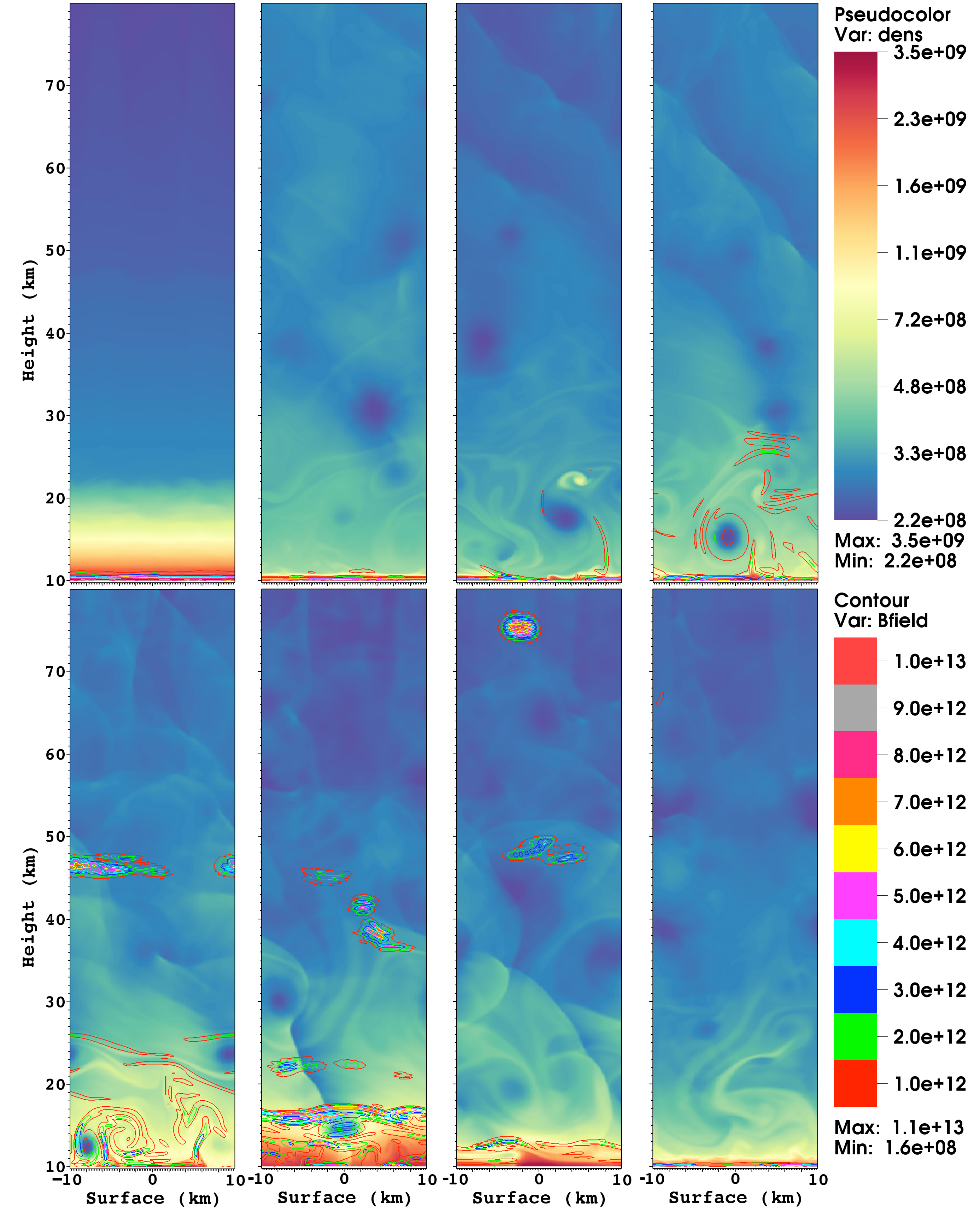}
   \caption{Magnetic field evolution in the post-hyperaccretion phase for $\dot m=10\: \dot m_{0}$. We show various time-steps of the dynamics near neutron star surface. In the Upper panels: (from left to right: $t=0,\: 1,\: 3,\: 5$ ms). In the Bottom panels: (from left to right: $t=8,\: 12,\: 16,\: 20$ ms). Note the hydromagnetic buoyancy effects in action.}
   \label{fig:5}
\end{figure*}

Figure \ref{fig:5} shows density color maps of the quasi-hydrostatic envelope, with magnetic field contours superimposed. In the upper panels, we show several timescales of the magnetic field evolution post-hyperaccretion phase (from left to right: $t=0,\: 1,\: 3,\: 5$ ms). Initially, the magnetic field remains confined into the crust, but eventually, the magnetic pressure begins to be dominant and small portions of the magnetic field are disconnected from the bulk of the magnetic field and carried away for buoyancy effects.  In the bottom panels, we present various timescales of the magnetic field evolution (from left to right: $t=8,\: 12,\: 16,\: 20$ ms), showing several MHD instabilities and buoyancy effects more notorious. Since the magnetic field no longer feels the strong accretion on it, we observe that magnetic tension generates small hydromagnetic instabilities, which in turn result in some convection in the flow. This process allows that more small portions of magnetic field to continue disconnected from the bulk magnetic field within the crust. The magnetic bubbles escape the computational domain without returning to it. A few ten milliseconds later,  the system finally achieves a transient relaxation regime. To where it was possible to follow the magnetic field evolution, there were not more drastic changes in the topology and the dynamics of the magnetic field. That is, although a significant portion of the magnetic field is lost by buoyancy effects, the bulk of the magnetic field remains confined into the thin NS crust.
Note that the transport of heat from the NS core to the stellar surface is determined by the thermal conductivity of the crust. Both thermal conductivity and electrical resistivity depend on several ingredients: the structure of the crust, the presence and number of crystalline defects and impurities and the nuclear composition. During some stages of the NS cooling, neutrino emission from the crust may significantly contribute to total neutrino losses from the stellar interior. Moreover, depending on the strong accretion rate, the submerged magnetic field could be pushed into the mantle, where it will be crystallized. Here, solid state physics is dominant and a detailed study of the magnetic field evolution is beyond the scope of this work.
\begin{table*}       
\begin{center}
\caption[]{\small\sf Densities and radii of the zones evolved in the hypercritical accretion episode}\label{table1}
\begin{minipage}{190mm}
\begin{tabular}{l|c|c|ll}
\hline \hline
 {\small Zones}  & {\small Density}        &  {\small Radii} & {\small Description} &\\
                         &  (${\rm g~cm}^{-3})$ & (cm) &\\\hline

{\small I. New crust of NS surface}      &  {\small $\rho_{e}(T,B)\,\,$}   & {\small $r_{ns}=10^{6}\,\,$}  &  {\scriptsize 1. A new crust with strong {\bf B} is formed by hypercritical phase}. \\
{\small ($ r\leq [r_{ns}+r_c]$)}                             & &{\small $r_{c}\simeq 10^{2.5}$} & {\scriptsize 2. Magnetic field in the range $10^{11}\leq {\bf B}\leq10^{13}\,{\rm G}$ could be submerged}. \\
& & & {\scriptsize 3. Photons and neutrinos confined are thermalized to a few MeV.}\\\hline
{\small II. Quasi-hydrostatic envelope}     & {\small $10^{2.9} \,\left(\frac{r_{s}}{r}\right)^{3}$}  &{\small  $r_{s}\simeq10^{8.9}$} &  {\scriptsize 1. Reverse shock induces hypercritical accretion onto the new NS surface }.   \\
{\small ($[r_{ns}+r_c]\leq r\leq r_s$)}                            & & & {\scriptsize 2. A new expansive shock is formed by the material accreted and bounced off}. \\
 & & & {\scriptsize 3. The high pressure close to NS surface allows the $e^{\pm}$ process to be dominant}.\\ \hline

{\small III. Free-fall}    &  {\small $10^{-1.24}\,\biggl(\frac{r}{r_h} \biggr)^{-3/2}$}       &{\small $r_h= 10^{10.8}$}     & {\scriptsize 1. Material begin falling with velocity $\sqrt{\frac{2GM}{r}}$ and density $\frac{\dot{m}}{4\pi r^2 v(r)}$}. \\
{\small ($r_s\leq r\leq r_h$)}                             & &    & \\\hline
 
{\small IV. External layers}    & {\small $10^{-5.2} A \left(\frac{R_\star}{r} -1\right)^{k} $}       & {\small $\,\,R_{\star}\simeq10^{12.5}$}  &{\scriptsize 1. The typical profile  of the external layers is presented}. \\
{\small ($r_h\leq r$)}                             &{\small $(k,A)=\cases{
{\small (2.1,20)};   r_h< r < r_a, \cr 
{\small (2.5,1)};     r>r_a.\cr
}
$}
&  {\small $r_a= 10^{11}$}      &\\
\hline\hline
\end{tabular}
\end{minipage}
\end{center}
\end{table*}
%
%
\section{Neutrinos}\label{sec-Neutrinos}
Neutrinos play an important role in the dynamics of supernovae. These particles not only provide an essential probe of the core collapse phenomena, but also probably play an active role in the catastrophic explosion mechanism.  Neutrinos play an active role in depositing energy in the region behind the outgoing shock wave and in the potential formation of heavy nuclei via r-process nucleosynthesis. Current numerical simulations show that this is the case (see,  \citep{2012MNRAS.426.1940K, 2016MNRAS.459.4174G}).\\
In the beginning of the core-collapse, when densities in the Fe-core rise above $\sim 10^9-10^{10}\text{ g cm}^{-3}$, electrons are captured by nuclei with electron neutrinos freely escaping from the collapsing star. When the core density becomes high enough to trap neutrinos, the resulting chemical equilibrium allows the diffusive emission of all flavors of neutrinos. Almost all of the gravitational binding energy will eventually be emitted in the form of neutrinos.\\
In the hypercritical phase, neutrino energy losses are dominated mainly by the annihilation process, which involves the formation of a neutrino-antineutrino pair when an electron-positron pair is annihilated near the stellar surface.  Table 3 summarizes the neutrino cooling processes, including the Cooper pairs formation of free neutrons (\textit{n}) for comparison purposes. The subscript \textit{x} indicate the neutrino flavor: electron, muon, or tau.
\subsection{Neutrino Oscillation}
Measurements of neutrino fluxes in solar, atmospheric and accelerator experiments have showed convincing evidences of neutrino oscillations.   The dynamics of the neutrino oscillations is solved by the Schrodinger equation {\small $i\frac{d\vec{\nu}}{dt}=H\vec{\nu}$}, where the state vector in the 3-dimensional flavor basis is {\small $\vec{\nu}\equiv(\nu_e,\nu_\mu,\nu_\tau)^T$}.  The Hamiltonian of this system is {\small $H=U\cdot H^d_0\cdot U^\dagger+\text{diag}(V_{\text{eff}},0,0)$} with {\small $H^d_0=\frac{1}{2E_\nu}\text{diag}(-\delta m^2_{21},0,\delta m^2_{32})$},  $V_{\text{eff}}$ is the neutrino effective potential and $U$ the three neutrino mixing matrix shown in  Ref. \cite{gon08}.    The  oscillation length of the transition probability  is given by  {\small $ l_{\text{osc}}=l_v/\sqrt{\cos^2 2\theta_{13} (1-\frac{2 E_{\nu} V_{\text{eff}}}{\delta m^2_{32} \cos 2\theta_{13}})^2+\sin^2 2\theta_{13}}$},  where $l_v=4\pi E_{\nu}/\delta m^2_{32}$ is the vacuum oscillation length and $E_\nu$ is the neutrino energy. The resonance condition is in the form
\be\label{reso3}
V_{\text{eff}}-5\times 10^{-7}\frac{\delta m^2_{32,\text{eV}}}{E_{\nu,\text{MeV}}}\,\cos2\theta_{13}=0\,.
\ee
In the case of 2-dimensional flavor bases \cite[e.g. see][]{2014MNRAS.437.2187F}, the  oscillation length is {\small $L_{osc}=L_v/\sqrt{\cos^2 2\theta (1-\frac{2 E_{\nu} V_{\text{eff}}}{\delta m^2 \cos 2\theta}
    )^2+\sin^2 2\theta}$} with $L_v=4\pi E_{\nu}/\delta m^2$, and  the resonance condition is
\be
V_{\text{eff}} -  \frac{\delta m^2}{2E_{\nu}} \cos 2\theta = 0\,.
\label{reso}
\ee
The three- and two-mixing (solar, atmospheric and accelerator; $\theta_{ij}$ and $\delta m_{ij}$) parameters are shown in \cite{aha11, abe11a, 1998PhRvL..81.1774, aha11,  wen10}. The neutrino effective potentials are given as follows.
\subsubsection{Effective Potential}
In the hypercritical accretion phase, thermal neutrinos are generated by the processes described in \cite{2008LRR....11...10C}.  These neutrinos going through the star oscillate in accordance with the density of each region  (see Table \ref{table1}). 

\paragraph{\bf Region I.}
On the NS surface,  the thermalized plasma is submerged in a magnetic field larger than  $>10^{10 }$ G. The neutrino effective potential for $1 \gg \frac{E^2_\nu}{m^2_W}$  is written as \citep{2014ApJ...787..140F, 2015MNRAS.451..455F}
\be\label{Veffm}
V_{\text{eff}}=V_0\,\sum^{\infty}_{l=0}(-1)^l\sinh\left\{\beta\mu(l+1)\right\}  \left[F-G\cos\varphi \right]\,, 
 \ee
where {\small $V_0=\frac{\sqrt2\,G_F\,m_e^3 B}{\pi^2\,B_c}$}, $\mu$ the chemical potential,  $\varphi$ is the angle between the neutrino momentum and the direction of magnetic field, $G_F$ is Fermi constant, $m_W$ is the W-boson mass and the functions  are 
\bary
F&=& K_1(\sigma_l)+2\sum^\infty_{n=1}\sqrt{1+2\,n\,\frac{B}{B_c}} K_1\left(\sigma_l \sqrt{1+2\,n\,\frac{B}{B_c}} \right)\,,\nonumber\\
G&=& K_1(\sigma_l)-2\sum^\infty_{n=1}\sqrt{1+2\,n\,\frac{B}{B_c}}\frac{E^2_\nu}{m^2_W} K_1\left(\sigma_l \sqrt{1+2\,n\,\frac{B}{B_c} }\right)\,,\nonumber\\
\eary
where K$_i$ is the modified Bessel function of integral order i and $\sigma_l=\beta m_e(l+1)$ with  $\beta=1/T$ and $m_e$ is the electron mass.  \\
\\
\paragraph{\bf Regions II, III and IV.}
In regions II, III and IV, neutrinos will undergo different effective potentials. It can be written as
\be
V_{\text{eff}}=\sqrt2 G_F\, N_A\,\rho(r)\,Y_e\,,
\ee
where $Y_e$ is the number of electron per nucleon,  $N_A$ is the Avogadro's number and $\rho(r)$ is the density given at different distances, as shown in Table \ref{table1}.\\
\subsection{Neutrino Expectation}
The number of neutrino events can be estimated in neutrino observatories as  Hyper-Kamiokande experiment. At 8 km south of its predecessor Super-Kamiokande (SK),  the Hyper-Kamiokande (HK) detector with $\sim$25 times larger than that of SK will be built as the Tochibora mine of the Kamioka Mining and Smelting Company, near Kamioka town in the Gifu Prefecture, Japan. Among the goals of this observatory are the astrophysical neutrino detections as well as and the studies of neutrino oscillation parameters \citep{2011arXiv1109.3262A, 2014arXiv1412.4673H}. The number of neutrinos to be detected in HK is 
\be
N_{ev}\simeq (3.4\times 10^{-9})\, T\, \int_{E'_{\bar{\nu}_e}}  \left( \frac{E_{\bar{\nu}_e}}{\text{MeV}}\right)^2  \frac{dN}{dE_{\bar{\nu}_e}}\,dE_{\bar{\nu}_e}\,,
\ee

\noindent where $T$ is the observed time of the neutrino burst and $dN/dE_{\bar{\nu}_e}$ is the neutrino spectrum.\\
\section{Conclusions}\label{sec-Results}

\begin{figure}
   \centering
   \includegraphics [width=0.75\textwidth]{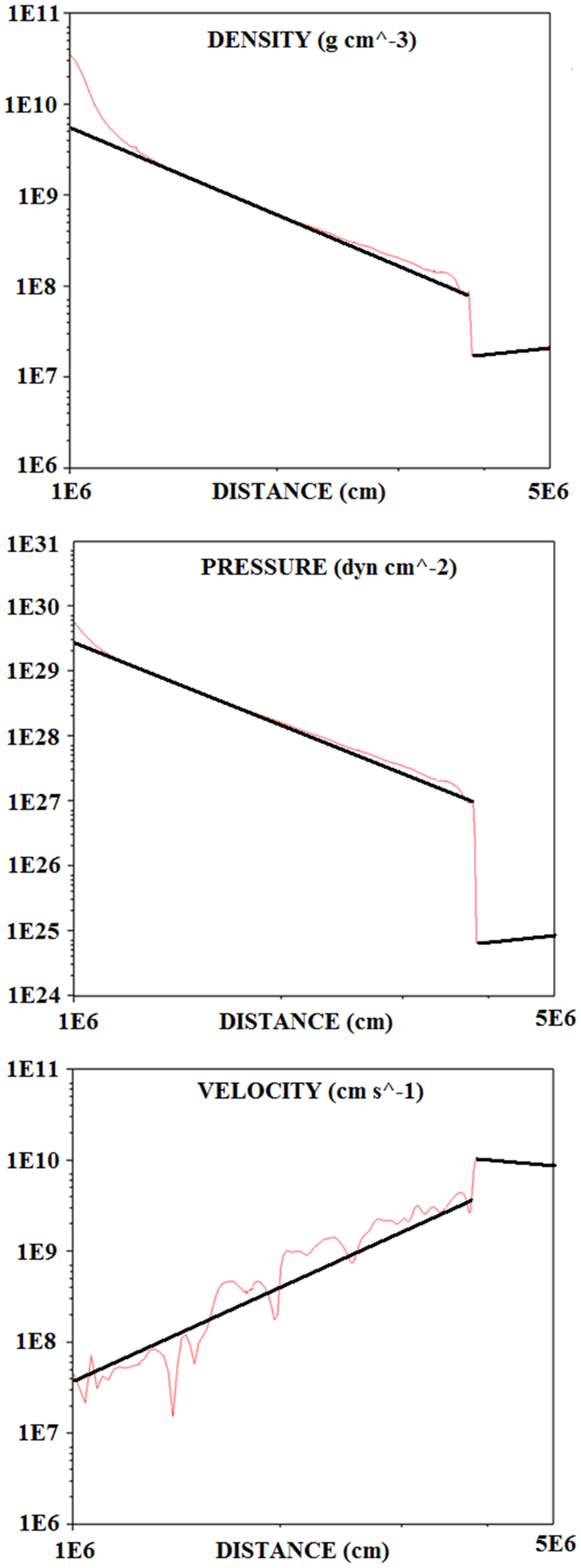}
   \caption{Profiles of density, pressure and velocity when the system reach a quasi-hydrostatic equilibrium. In red are the profiles obtained with the numerical simulations and superimposed, in black, are the analytical profiles from \citep{Chevalier2005}.}
   \label{fig:3}
\end{figure}

\begin{figure*}
\centering
\includegraphics[width=0.6\textwidth]{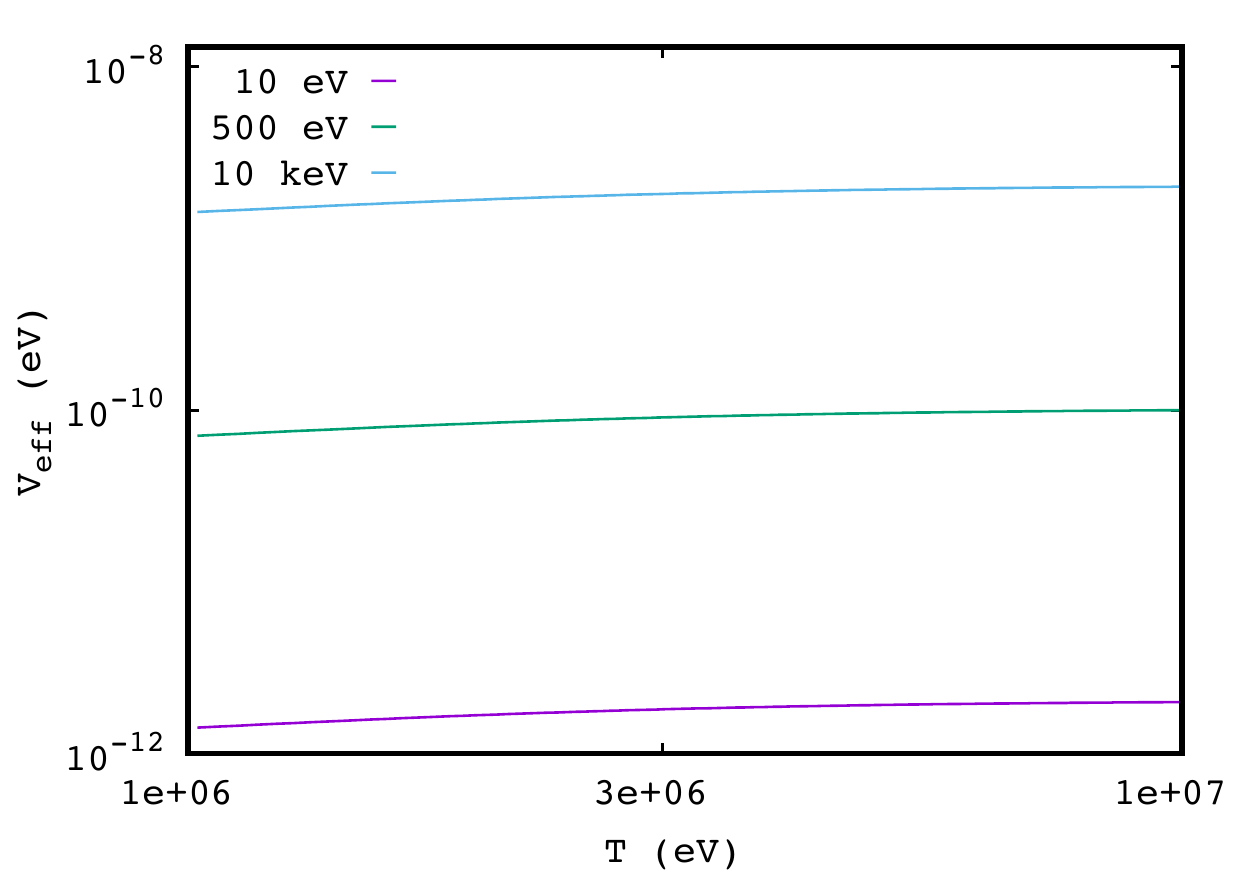}
\caption{Neutrino effective potential as a function of temperature and chemical potential for a magnetic field of $2.9\times10^{10}$ G. }
\label{Veff}
\end{figure*}

\begin{figure*}
\centering
\includegraphics[width=0.84\textwidth]{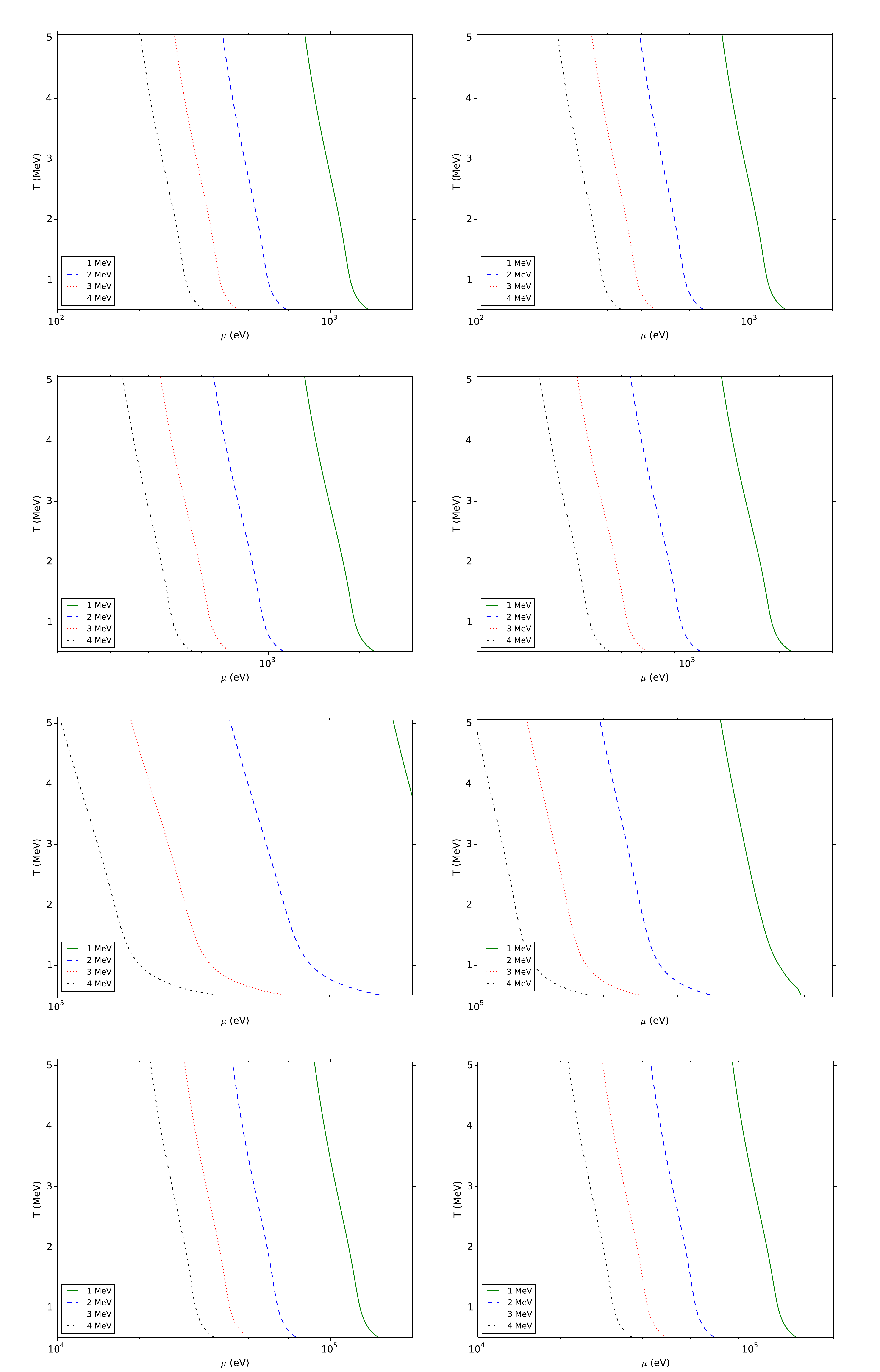}
\caption{Temperature vs  chemical potential  for which the resonance condition is satisfied.  Several neutrino energies (E$_\nu$=1, 2, 3 and 4 MeV), angles ($\varphi=0^\circ$; the left-hand  column and $\varphi=90^\circ$; the right-hand  column) and the best fit values of neutrino mixing (from top to bottom: solar,  atmospheric, accelerator and three flavors) have been considered.}
\label{Rcond}
\end{figure*}

We have studied the dynamics of hyper-accretion onto the newly born NS in the CCO scenario. To do this, numerical 2D-MHD simulations have been performed using the AMR \textsc{flash} method. We found that for the hyper-accretion rates considered here, initially, the magnetic field resists to be submerged in the new crust suffering several episodes of magnetic reconnections and later, it returns to its original shape. This is very interesting because it allows us to analyze how the magnetic reconnection processes work in extreme  temperature conditions, density and strong magnetic fields.  Finally, the bulk of the magnetic field is submerged into the new crust by the hyper-accretion, with a quasi-hydrostatic envelope around it.  In addition, we have estimated the neutrino luminosity for all the involved processes which is mandatory for the analysis of propagation/oscillations through the regions of the pre-supernova.    We have observed that several peaks in the luminosity curve are very notorious. The same behavior is present in the curve of magnetic energy density. This corresponds to several episodes of magnetic reconnections driven by magnetic Rayleigh-Taylor instabilities. The code allows that the rich morphology and the main characteristics of such physical processes to be captured in detail.  Of course, the question of how the flow approaches the steady state in a time-depending situation remains as an open problem.  As noted by Chevalier \cite{Chevalier2005}, the study of quasi-hydrostatic envelopes around NS is like a runaway process in which the neutrino losses lead to the gravitational contraction of the envelope. It  increases the pressure at the base of the envelope and then the neutrino losses.  While neutrino losses play an important role in the envelope structure, the optical depth of photons is large enough so that probably the radiative transfer only plays a role in the later evolution. The escape of radiation at the Eddington limit can have a dramatic effect on the evolution of the hyper-accretion process. A detailed study of the thermal structure of the gas surrounding the NS is needed to determine whether it occurs and also to analyze if the quasi-hydrostatic envelope is evaporated by neutrinos, photons or other processes. We have verified that the steady state solutions give the most plausible structure for the NS envelopes under influence of the neutrino cooling near the newborn NS surface.\\
On the other hand, when the hyper-accretion phase is over, the magnetic field can suffer a growing episode until it appears as a delayed pulsar for decades or centuries. Nevertheless, it is possible that large portions of the confined magnetic field may be pushed into the NS by the hyper-accretion, and it may be crystallized in the mantle. The presence of a crystal lattice of atomic nuclei in the crust is mandatory for the modeling of the subsequent radio-pulsar glitches. 
These processes, the eventual growth of the bulk magnetic field post-hyperaccretion and the magnetic field crystallization processes inside NS, require more detailed studies that are out of the scope of this paper. However, numerical simulations with more refinement, more physical ingredients and more degrees of freedom in the system, offer new and unique insight into this problem.  Although further numerical studies of these phenomena are necessary, the results of this work can give us a glimpse of the complex dynamics around the newly born NS, moments after the core-collapse supernova explosion. \\
Using the quantities observed such as distance $2.2\pm 0.3$ kpc \citep{1995AJ....110..318R}, neutrino Luminosity  $L_{\bar{\nu}_e}=(2.6\pm 0.2)\times10^{48}\:\mathrm{erg\, s^{-1}}$ (see Figure \ref{fig:3}), effective volume $V\simeq 0.56\times 10^{12}\,{\rm cm^3}$ \citep{2014arXiv1412.4673H}  and the average neutrino energy  $<E_{\bar{\nu}_e}>\simeq\, 4\, {\rm  MeV}$,    the number of events expected from the hypercritical phase is 1323$\pm$648.   Following Refs. \cite{2004mnpa.book.....M} and  \cite{1989neas.book.....B},  we estimate the number of the initial neutrino burst from the NS formation. Taking into consideration  the duration of the neutrino burst $t\simeq10$ s and T$\approx$ 4 MeV  \citep{2007fnpa.book.....G}, the average neutrino energy and the total fluence equivalent for 1E 1207.4-5209 becomes $<E_{\bar{\nu}_e}>\simeq  13.5\pm 3.2\, {\rm  MeV}$ and  $\Phi\approx(1.30\pm 0.54) \times 10^{12}\, \bar{\nu}_e \,{\rm cm^{-2}}$, respectively.  Therefore,  the total number of neutrinos released and the total radiated luminosity during the NS formation are $N_{tot}=6\,\Phi\, 4\pi\, d_z^2\approx(4.55\pm 1.82) \times 10^{57}$ and  $L_\nu\approx \frac{N_{tot}}{t}\times <E_{\bar{\nu}_e}>\approx (9.84 \pm3.94) \times 10^{51}$ erg/s, respectively. Finally, the number of events expected during the NS formation on HK experiment is $N_{ev}\simeq (4.82\pm 1.91)\times 10^6$. Therefore,  during the NS formation  $(3.65\pm 1.46)\times 10^3$ events are expected  more than the  hyper-accretion phase.  Taking into consideration the observable quantities given in Table \ref{t1},  the same calculation for other CCOs are reported in Table \ref{Table:obs}.\\
Thermal neutrinos created during the hypercritical accretion phase will oscillate in matter due to electron density in regions I, II, III and IV, and in vacuum into Earth.   In the new crust of NS surface (region I), the plasma  thermalized at $\simeq$ 4 MeV is submerged in a magnetic field of $\simeq$ 2.6$\times 10^{10}$ G. Regarding the neutrino effective potential given in this region,  the neutrino effective potential as a function of temperature  and chemical potential   for B$\simeq$ 2.6$\times 10^{10}$ G is plotted, as shown in Figure \ref{Veff}.  This figure shows  the positivity of the effective potential ($V_{\text{eff}}>$ 0), hence neutrinos can oscillate resonantly.   Using the two and three- neutrino mixing parameters we study  the resonance condition, as shown in Figure \ref{Rcond}.   \citet{2014MNRAS.442..239F} calculated  the survival and conversion probabilities for the active-active ($\nu_{e,\mu,\tau} \leftrightarrow \nu_{e,\mu,\tau}$) neutrino oscillations in regions II, III and IV,  showing that neutrinos can oscillate resonantly in these regions due to  the density profiles of the collapsing material surrounding the progenitor.     Taking into consideration the oscillation probabilities in each region and in the vacuum,  the flavor ratios expected on Earth for neutrino energies of $E_{\nu}=$1, 5, 10 and 20 MeV were estimated.   Table \ref{Table:ratio} displays a small deviation from the standard ratio flavor 1:1:1.   In this calculation we take into account the neutrino cooling processes shown in \cite{2008LRR....11...10C}.   It is worth noting that our calculations of resonant oscillations were performed for neutrinos instead of anti-neutrinos, due to the positivity of the neutrino effective potential.\\
We conclude that the neutrino burst could be the only viable observable that can provide relevant evidence of  the hypercritical phase and therefore, the hidden magnetic field mechanism as the most favorable scenario to interpret the anomalous low magnetic fields estimated for CCOs.
\begin{table*}[h!]     
\begin{center}
\caption[]{Number of events expected from the hypercritical phase and NS formation.}\label{Table:obs}
\begin{minipage}{156mm}
\begin{tabular}{ l c c }
\hline\hline
{\small CCO}                   & {\small Events (hypercritical phase)}          &  {\small Events (NS formation)}  \\ 
& &{\small $\times10^6$}\\\hline\hline
{\small XMMU J173203.3-344518} &  {\small $625\pm300$}   & {\small $2.28\pm0.92$}     \\
{\small 1E 1207.4-5209}        & {\small $1323\pm648$}  & {\small $4.82\pm1.91$}   \\
{\small CXOU J160103-513353}   & {\small $256\pm129$} & {\small $0.93\pm0.38$}  \\
{\small 1WGA J1713.4-3949}     &   {\small $3789\pm1819$} &  {\small $13.80\pm5.52$}  \\
{\small XMMU J172054.5-372652} &  {\small $316\pm156$}   &   {\small $1.15\pm0.46$} \\
{\small CXOU J085201.4-461753} &   {\small $6403\pm3073$}   &   {\small $23.33\pm9.32$} \\
                      &                &          \\\hline\hline\end{tabular}
\end{minipage}
\end{center}
\end{table*}

\begin{table*}[h!]     
\begin{center}
\caption[]{The neutrino flavor ratio expected on Earth  for  $E_{\nu}=$ 1, 5, 10 and 20 MeV.}\label{Table:ratio}
\begin{minipage}{156mm}
\begin{tabular}{lccccccccc}
\hline
 
 $E_{\nu}$  & XMMU J173203.3 & 1E 1207.4 & CXOU J160103 & 1WGA J1713.4 & XMMU J172054.5 & CXOU J085201.4 \\\hline \hline 

{\small 1}     &  {\small 1.036:0.988:0.976}  & {\small 1.035:0.989:0.976}  & {\small 1.034:0.988:0.978}  & {\small 1.036:0.987:0.975}  &  {\small 1.037:0.987:0.976} &  {\small 1.034:0.989:0.977} \\\hline

{\small 5}   & {\small 1.037:0.987:0.975}   &  {\small 1.038:0.986:0.975} &  {\small 1.035:0.989:0.975}&  {\small 1.036:0.987:0.976}  &  {\small 1.037:0.988:0.974} &  {\small 1.035:0.988:0.976} \\\hline

{\small 10}  & {\small 1.040:0.977:0.984} &  {\small 1.041:0.976:0.984} &  {\small 1.042:0.975:0.984} &  {\small 1.039:0.978:0.985}  &  {\small 1.040:0.978:0.983} &  {\small 1.041:0.977:0.983}  \\\hline

{\small 20}  & {\small 1.045:0.933:0.982} &    {\small 1.044:0.934:0.982}  &   {\small 1.043:0.934:0.983} &   {\small 1.044:0.933:0.983}  &  {\small 1.046:0.932:0.982} &  {\small 1.043:0.934:0.983} \\\hline
 
\end{tabular}
\end{minipage}
\end{center}
\end{table*}


%
\acknowledgements
We thank  Dany Page Rollinet and John Beacom  for useful discussions. NF acknowledge financial  support  from UNAM-DGAPA-PAPIIT  through  grant  IA102917. The software used in this work was in part developed by the DOE NNSA-ASC OASCR FLASH Center at the University of Chicago. 
%
%
%
%

%


%
\clearpage
%
%
%
%
%

\end{document}